\documentclass[useAMS,usenatbib]{mn2e}
\usepackage{graphicx}
\usepackage{url}
\usepackage[utf8]{inputenc}
\usepackage{fancyref}
\usepackage{hyperref}
\hypersetup{colorlinks,breaklinks,
            urlcolor=[rgb]{0.2,0.2,0.2},  
            linkcolor=[rgb]{0,0,0},  
            citecolor=[rgb]{0,0,1.0}}  

\newcommand{\teff}{${T}_{\mathrm{eff}}$}
\newcommand{\logg}{$\log{g}$}
\newcommand{\msun}{$M_{\odot}$}
\newcommand{\mstar}{$M_{\star}$}
\newcommand{\rsun}{$R_{\odot}$}

\newcommand{\kms}{km s$^{-1}$}
\newcommand{\muhz}{$\mu$Hz}

\newcommand{\tar}{J1136+0409}
\newcommand{\bjd}{BJD(TDB)}
\newcommand{\Ion}[2]{#1{\,\scriptsize #2}}

\def\aap{A\&A}
\def\apjl{ApJ}
\def\apj{ApJ}

\def\mnras{MNRAS}
\def\nat{Nature}
\def\pasp{PASP}

\def\araa{ARA\&A}

\title[Internal effects of common-envelope evolution]{Insights into internal effects of common-envelope evolution using the extended {\em Kepler} mission}

\author[Hermes et al.]{J.~J.~Hermes,$^{1}$\thanks{j.j.hermes@warwick.ac.uk}
B.~T.~G\"{a}nsicke,$^{1}$
A.~Bischoff-Kim,$^{2}$
Steven~D.~Kawaler,$^{3}$
\newauthor
J.~T.~Fuchs,$^{4}$
B.~H.~Dunlap,$^{4}$
J.~C.~Clemens,$^{4}$
M.~H.~Montgomery,$^{5}$
P.~Chote,$^{1}$
\newauthor
Thomas~Barclay,$^{6,7}$
T.~R.~Marsh,$^{1}$
A.~Gianninas,$^{8}$
D.~Koester,$^{9}$
D.~E.~Winget,$^{5}$
\newauthor
D.~J.~Armstrong,$^{1}$
A.~Rebassa-Mansergas,$^{10}$ and
M.~R.~Schreiber$^{11}$
\\
$^{1}$Department of Physics, University of Warwick, Coventry\,-\,CV4~7AL, UK\\
$^{2}$Penn State Worthington Scranton, Dunmore, PA\,-\,18512, USA\\
$^{3}$Department of Physics and Astronomy, Iowa State University, Ames, IA\,-\,50011, USA\\
$^{4}$Department of Physics and Astronomy, University of North Carolina, Chapel Hill, NC\,-\,27599-3255, USA\\
$^{5}$Department of Astronomy, University of Texas at Austin, Austin, TX\,-\,78712, USA\\
$^{6}$NASA Ames Research Center, Moffett Field, CA\,-\,94035, USA\\
$^{7}$Bay Area Environmental Research Institute, 596 1st Street West, Sonoma, CA\,-\,95476, USA\\
$^{8}$Homer L. Dodge Department of Physics and Astronomy, University of Oklahoma, 440 W. Brooks St., Norman, OK\,-\,73019, USA\\
$^{9}$Institut f\"{u}r Theoretische Physik und Astrophysik, University of Kiel, Kiel\,-\,D-24098, Germany\\
$^{10}$Kavli Institute for Astronomy and Astrophysics, Peking University, Beijing\,-\,100871, China\\
$^{11}$Departamento de F\'isica y Astronom\'ia, Universidad de Valpara\'iso, Avenida Gran Breta\~na 1111, Valpara\'iso, Chile
}

\begin{document}

\maketitle

\label{firstpage}

\begin{abstract}

We present an analysis of the binary and physical parameters of a unique pulsating white dwarf with a main-sequence companion, SDSS~J1136+0409, observed for more than $77$\,d during the first pointing of the extended {\em Kepler} mission: {\em K2} Campaign 1. Using new ground-based spectroscopy, we show that this post-common-envelope binary has an orbital period of 6.89760103(60)\,hr, which is also seen in the photometry as a result of Doppler beaming and ellipsoidal variations of the secondary. We spectroscopically refine the temperature of the white dwarf to 12\,330$\pm$260\,K and its mass to 0.601$\pm$0.036\,\msun. We detect seven independent pulsation modes in the {\em K2} light curve. A preliminary asteroseismic solution is in reasonable agreement with the spectroscopic atmospheric parameters. Three of the pulsation modes are clearly rotationally split multiplets, which we use to demonstrate that the white dwarf is not synchronously rotating with the orbital period but has a rotation period of $2.49\pm0.53$\,hr. This is faster than any known isolated white dwarf, but slower than almost all white dwarfs measured in non-magnetic cataclysmic variables, the likely future state of this binary.

\end{abstract}


\section{Introduction}
\label{sec:intro}

\begin{figure*}
\centering{\includegraphics[width=0.92\textwidth]{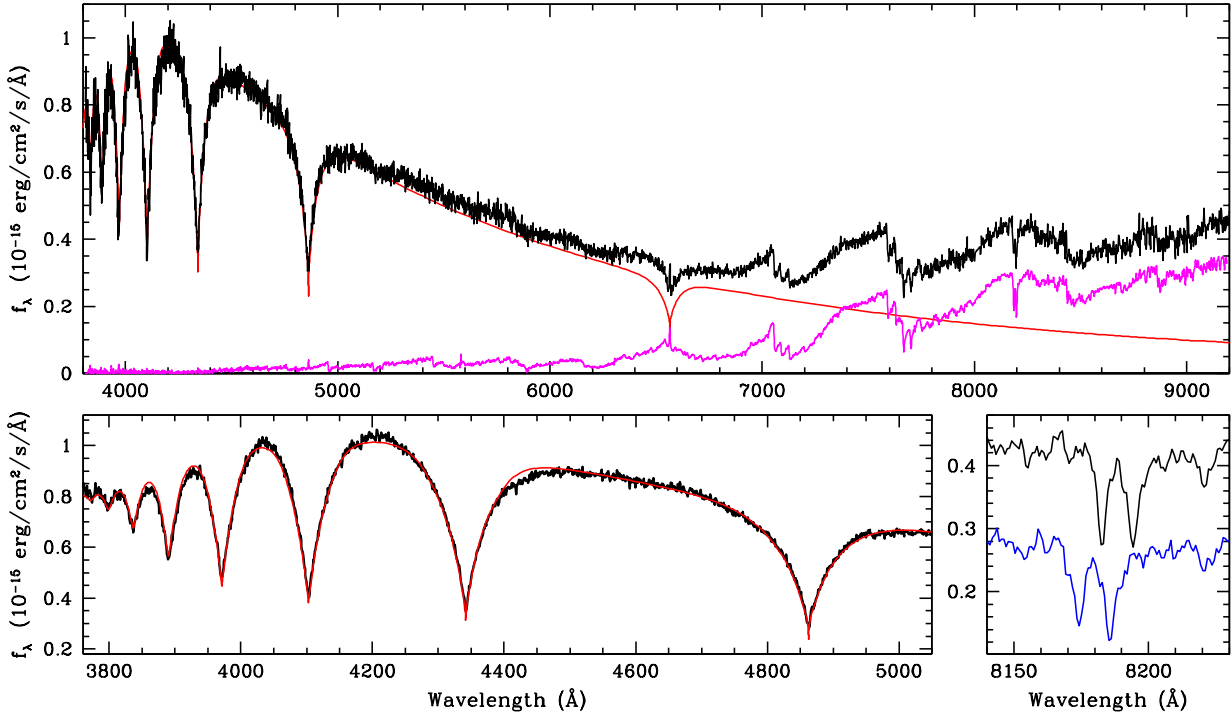}}
\caption{Spectroscopic observations of the WD+dM binary \tar. The top panel shows the discovery spectrum from SDSS. Underplotted in red is the best-fitting SDSS white dwarf model, as well as an M6 template in magenta estimating the contribution of the M-dwarf companion at redder wavelengths. The bottom left panel details our high signal-to-noise SOAR spectrum, from which we refine the white dwarf atmospheric parameters (red line, see Section~\ref{sec:spec}). The bottom right panel shows two different epochs of our VLT/FORS2 observations to calculate the radial velocity of the M-dwarf companion using the \Ion{Na}{I} doublet (see Section~\ref{sec:bin:rv}). \label{fig:spec}}
\end{figure*}

White dwarfs are the endpoint for all stars less massive than about $8-10$\,\msun, making them the fossils of more than 97\% of Galactic stellar content \citep{Williams09,Doherty15}. They are representative of the general population of all stars, providing benchmarks that can calibrate theories of both stellar and binary evolution. 

White dwarfs do not sustain substantial nuclear burning, and their evolution is almost exclusively defined by passive cooling of the residual thermal energy of ions in the core. Fortunately, when hydrogen-atmosphere (DA) white dwarfs reach an appropriate effective temperature (roughly $12,500$\,K), their non-degenerate atmosphere develops a partial ionization zone, impeding energy transport and driving global pulsations; these are the ZZ Ceti stars (or so-called DAVs). Asteroseismology, enabled by matching the periods of observed luminosity variations to well-calibrated theoretical models, provides unparalleled insight into the interiors of these white dwarfs (see reviews by \citealt{WinKep08}; \citealt{FontBrass08}; and \citealt{Althaus10}).

The development of a surface convection zone and the subsequent driving of pulsations are believed to be a naturally occurring phase for all white dwarfs (e.g., \citealt{Bergeron04, Castanheira07}). This so far holds regardless of overall stellar mass \citep{Giovannini98} or core composition \citep{Hermes13}.

White dwarfs found in binaries should also pulsate when they reach the appropriate temperature, and recently the first pulsating white dwarfs with main-sequence companions were discovered \citep{Pyrzas15}. Roughly two-thirds of the known white dwarfs with M dwarf companions are wide enough that the progenitors evolved as if they were single stars (e.g., \citealt{NGM11}). The others evolved through a common envelope with a close companion and are post-common-envelope binaries (PCEBs).

Most PCEBs are the detached precursors to cataclysmic variable systems. Nearly all progenitors of observed Supernovae~Ia, independent of single vs. double-degenerate channel, are believed to go through at least one common-envelope phase. Pulsating white dwarfs in PCEBs thus hold the exciting potential to allow for the first empirical test of the effects binary interaction has on the remnant white dwarf internal structure and chemical profiles, and can possibly directly constrain Supernovae~Ia boundary conditions.

Serendipitously, the only pulsating white dwarf in a confirmed PCEB discovered by \citet{Pyrzas15}, SDSS J113655.17+040952.6 (hereafter \tar), was observable during the first science field of the two-wheel-controlled {\em Kepler} mission ({\em K2} Campaign 1). The extended {\em Kepler} mission delivers space-based, time-series photometry on targets for up to 85\,d along the ecliptic (see \citealt{Howell14}).

\tar\ was spectroscopically identified as a composite WD+dM from a Sloan Digital Sky Survey (SDSS) spectrum. Model-atmosphere fits to the Balmer absorption lines initially suggested the white dwarf had an effective temperature of $11\,700\pm150$\,K and \logg\ $= 7.99\pm0.08$ \citep{Rebassa-Mansergas12a}, placing it in the empirical DAV instability strip. Pulsations were confirmed by a short (1.0\,hr) run with the ULTRACAM instrument mounted on the 3.5m New Technology Telescope, with significant periods near 276.5\,s and 182.2\,s \citep{Pyrzas15}.
 
For nearly 78\,d in 2014~August, the {\em Kepler} spacecraft observed \tar\ every minute, delivering an exceptional light curve of the first known pulsating white dwarf in a PCEB. We present here a detailed analysis of the pulsations and binary parameters of \tar, enabled by space-based photometry, as well as spectroscopy from the 8.2-m Very Large Telescope (VLT) and the 4.1-m Southern Astrophysical Research (SOAR) telescope.


\begin{figure*}
\centering{\includegraphics[width=0.85\textwidth]{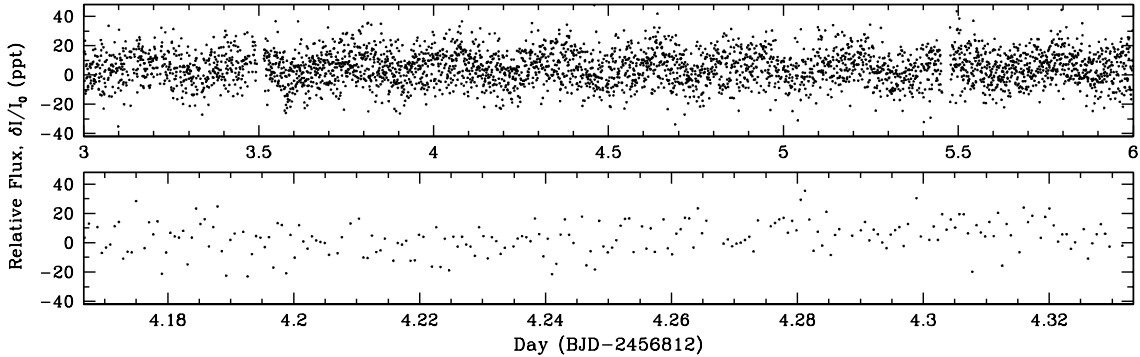}}
\caption{A portion of the raw {\em K2} Field 1 light curve obtained for \tar. This minute-cadence photometry has been extracted from the pixel-level {\em K2} data and de-trended to correct for attitude tweaks in the telescope pointing, and has been 4$\sigma$ clipped. The top panel shows 3\,d (less than 4\,per\,cent of our full dataset) and is dominated by variability at the orbital period of 6.9\,hr (see Section~\ref{sec:bin:lc}). The bottom panel shows a 4\,hr stretch. The white dwarf pulsations are not visible above the point-to-point scatter. 1 ppt = 0.1\,per\,cent.
\label{fig:lc}}
\end{figure*}

\section{Observations and Reductions}
\label{sec:obs}

\subsection{Spectroscopic Observations}
\label{sec:obs:spec}

The identification of \tar\ as both a candidate pulsating white dwarf and a close WD+dM binary arose from an ancillary SDSS science fiber placed on the target. That spectrum, visible in the top panel of Fig.~\ref{fig:spec}, clearly displays a composite of a white dwarf at bluer wavelengths and an M dwarf dominating at redder wavelengths. Fits for the white dwarf temperature put it within the empirical DAV instability strip, and the SDSS sub-spectra showed significant radial-velocity variability over just 18\,min, suggesting it is a close binary system. To better constrain both unique aspects of the system, we obtained follow-up spectroscopy.

In order to solve the radial velocity of the dM component of the system, we obtained 11 service-mode observations using the FORS2 instrument on the 8.2\,m VLT/UT1 (Antu) between 2009~April~4 and 2014~May~27. These intermediate-resolution spectra were all taken using the 1028z grism and a 1\arcsec\ slit, covering a spectral region of $7830-9570$\,\AA\ with a dispersion of roughly 0.8\,\AA\ per pixel and 2.4\,\AA\ resolution. We were exclusively interested in the sharp \Ion{Na}{I} absorption doublet at rest wavelengths of 8183.27 and 8194.81 \AA\ (e.g., \citealt{Rebassa-Mansergas12b}). Further details describing our reduction and analysis methods for the VLT data can be found in \citet{Schreiber08}.

We display in the bottom right panel of Fig.~\ref{fig:spec} two epochs of the \Ion{Na}{I} doublet, separated by less than 2.0\,hr on 2014~May~27. The analysis of these radial velocities and a full table of the observing epochs and exposure times are detailed in Section~\ref{sec:bin:rv}. 

Additionally, we sought to improve the white dwarf atmospheric parameters, and obtained higher signal-to-noise (S/N) spectra using the Goodman spectrograph on the 4.1\,m SOAR telescope \citep{Clemens04}. These blue spectra, 4$\times$420\,s exposures taken consecutively on 2015~January~29, cover the higher-order Balmer lines and range from roughly $3600-5300$\,\AA\ with a dispersion of roughly 0.8\,\AA\ per pixel and 2.3\,\AA\ resolution. We used a 3\arcsec\ slit, and the seeing was just under 1\arcsec. Each exposure has a S/N$\sim70$ per resolution element in the continuum at 4600 \AA.

For all of our follow-up spectroscopy, we processed the images using the {\sc starlink} packages {\sc figaro} and {\sc kappa}. The spectra were optimally extracted \citep{Horne86} using the {\sc pamela} package \citep{Marsh89}, and wavelength calibrated using the many sky emission lines present. We subsequently flux calibrate and rebinned the spectra to a heliocentric frame using the {\sc molly} package\footnote{\href{http://www.warwick.ac.uk/go/trmarsh}{http://www.warwick.ac.uk/go/trmarsh}}. We have estimated the spectral resolution from measurements of the full widths at half maximum of arc-lamp and night-sky emission lines.

\subsection{Space-Based Photometry}
\label{sec:obs:K2}

The {\em Kepler} spacecraft provides an unprecedented capability for acquiring uninterrupted times series of faint variable stars, and the photometry of \tar\ (EPIC 201730811, $K_p=17.2$\,mag) opens a new era in seismology of white dwarfs in binary systems. Our short-cadence exposures, observed every 58.8\,s, span nearly 78 days, from 2014~June~4 03:02:00 UT to 2014~August~20 19:55:38 UT.

With only two functional reaction wheels, the pointing of the {\em Kepler} spacecraft must be corrected periodically (with checks every 6\,hr) by firing its thrusters. Thruster firings create significant discontinuities in the photometry and must be detrended. We extracted the light curve using a modified version of the {\sc kepsff} code based on \citet{Vanderburg14} within the {\sc pyke} package maintained by the {\em Kepler} Guest Observer office. We have also applied an algorithm to select the pixel mask that looks at contiguous pixels above the background; our method is described in more detail in \citet{HermesGD1212}. A portion of the light curve is displayed in Fig.~\ref{fig:lc}.

Because the star has variability with a period near to 6\,hr, the thruster-firing period, we have utilized an independent extraction of the data using the tools outlined in \citet{Armstrong15} to detrend the light curve. However, the r.m.s. scatter of this extraction was slightly higher, so we do not adopt it for our full analysis.

We have removed all points with poor quality flags, as well as those affected by thruster firings, removing 32\,265 points. We subsequently clipped all points falling more than 4$\times1.4826$ times the median absolute deviation from the median of the light curve, removing 351 points and ultimately leaving us with 91\,144 exposures over 77.8724\,d. We have also excluded a brightening event 95.1\,min in duration beginning at 2456844.36916 \bjd; we have inspected the raw photometry using the {\sc k2flix} package\footnote{\href{http://barentsen.github.io/k2flix/}{http://barentsen.github.io/k2flix/}} \citep{Barentsen15} and shown that this brightening corresponds to a solar system body passing through our target aperture. Pointing into the ecliptic, {\em K2} encounters many asteroids as they pass through its field of view.

Our final {\em K2} light curve has a duty cycle of 79.7\,per\,cent and has a formal frequency resolution of 0.149\,\muhz. The point-to-point scatter is evident by eye in Fig.~\ref{fig:lc} (exposures have a mean uncertainty of 13.0\,ppt), and unfortunately overwhelms the ability to detect the pulsations by eye. Given the long baseline, the median level in a 1000-\muhz-wide, empty region of the Fourier transform centered at 1500\,\muhz\ is roughly 0.064\,ppt (64\,ppm). We describe amplitudes in this manuscript in units of parts per thousand (ppt), where 1 ppt = 0.1\,per\,cent relative amplitude.

\section{Atmospheric Parameters}
\label{sec:spec}

From the original serendipitous SDSS spectrum of \tar, \citet{Rebassa-Mansergas12a} fitted the Balmer lines of H$\beta-$H$\epsilon$ following the method of \citet{Rebassa-Mansergas07}. They found for the white dwarf a \teff\ $=11\,700\pm150$\,K and \logg\ $= 7.99\pm0.08$.

We obtained higher-S/N SOAR spectroscopy to refine the atmospheric parameters of the pulsating component of \tar. White dwarf atmospheric parameters are most sensitive to the higher-order Balmer lines, especially the surface gravity (e.g., \citealt{Kepler06}), so we have used the blue-efficient Goodman spectrograph on the SOAR telescope (see Section~\ref{sec:obs:spec}).

We have fitted the six Balmer lines H$\beta-$H9 using the synthetic spectra computed by two independent groups. The first set of fits use the pure hydrogen atmosphere models and fitting procedure described in \citet{Gianninas11} and references therein, which employ the ML2/$\alpha = 0.8$ prescription of the mixing-length theory \citep{Tremblay10}. We have fitted each of the four individual SOAR spectra, and found the weighted mean of the atmospheric parameters to be \teff\ $=12\,680\pm150$\,K and \logg\ $= 7.97\pm0.03$. In addition, we have individually fit the same spectra using the pure hydrogen atmosphere models detailed in \citet{Koester10}, which also employ ML2/$\alpha = 0.8$. From these fits we find a weighted mean of the atmospheric parameters to be \teff\ $=12\,310\pm150$\,K and \logg\ $= 8.05\pm0.03$.

To accommodate the unknown systematic uncertainties, we have chosen to adopt the average of these two independent weighted mean fits, using the standard deviation of the two as the uncertainties: \teff\ $=12\,490\pm260$\,K and \logg\ $= 8.01\pm0.06$. These values represent the effective temperature and surface gravity from fits to models with a one-dimensional treatment of convection. The SDSS fits used ML2/$\alpha = 0.6$, which explains the disagreement with our new determination.

Our adopted atmospheric parameters require a correction to account for the higher-dimensional dependence of convection \citep{Tremblay13}. Thus, our final best-fitting determinations are \teff\ $=12\,330\pm260$\,K and \logg\ $= 7.99\pm0.06$ which corresponds to an overall white dwarf mass of $0.601\pm0.036$\,\msun\ and a radius of $R_{\rm WD}=0.013$\,\rsun\ using the mass-radius relations of \citet{Fontaine01}.

By measuring the secondary contribution to the SDSS spectrum (see the top panel in Fig.~\ref{fig:spec}), we estimate that the main-sequence component to \tar\ is of spectral type M$6\pm1$V. Using the empirical mass-radius relation of \citet{Rebassa-Mansergas07}, we estimate $M_{\rm sec} = 0.196\pm0.085$\,\msun\ and $R_{\rm sec} = 0.195\pm0.090$\,\rsun. From the flux-scaling factors of each component, we can estimate the overall distance to the system (e.g., \citealt{Rebassa-Mansergas12a}). From the white dwarf we find $d_{\rm WD} = 125\pm7$\,pc, and from the M dwarf we find $d_{\rm sec} = 121\pm56$\,pc, in good agreement.

\section{Binary Analysis}
\label{sec:bin}

\subsection{Secondary Component Radial Velocities}
\label{sec:bin:rv}

\begin{figure}
\centering{\includegraphics[width=0.995\columnwidth]{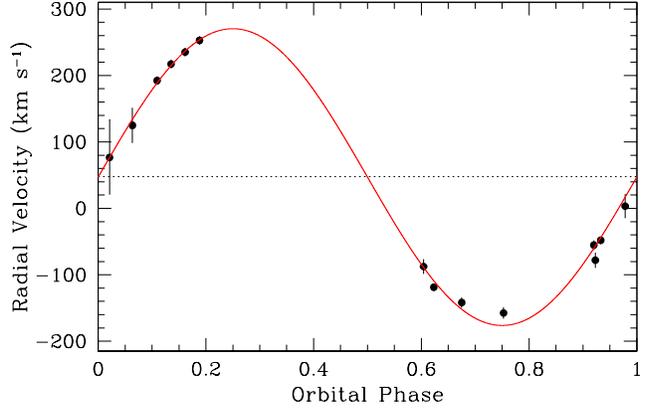}}
\caption{Radial velocity curve of the M dwarf companion in \tar, calculated from the \Ion{Na}{I} doublet. This curve is folded on the orbital period of 6.8976010\,hr, and features a semi-amplitude of $K_{\rm sec} = 222.6\pm3.8$ \kms. \label{fig:rv}}
\end{figure}

\begin{table}
\begin{center}
\vspace{-0.1in}
 \centering
  \caption{Journal of radial velocities from the \Ion{Na}{I} doublet.}\label{tab:rvjour}
  \begin{tabular}{@{}lrr@{}}
  \hline
BJD$_{\rm TDB}$ & Radial Velocity & Notes \\
  & (\kms) &  \\
 \hline
 2452642.987095  &  $76.6 \pm 56.1$ &  900s, SDSS subspec \\
 2452642.974837  &  $3.3 \pm 18.0$ &  900s, SDSS subspec \\
 2452642.999318  &  $124.9 \pm 26.5$ &  900s, SDSS subspec \\
 2454925.691559  &  $-118.7 \pm 5.5$  &  900s, VLT/FORS2 \\
 2454937.564039  &  $-48.0 \pm 5.7$  &  900s, VLT/FORS2 \\
 2456735.731791  &  $-87.6 \pm 11.0$ &  570s, VLT/FORS2 \\
 2456745.595020  &  $-78.0 \pm 11.6$ &  570s, VLT/FORS2 \\
 2456803.578535  &  $-141.8 \pm 7.7$  &  570s, VLT/FORS2 \\
 2456803.600854  &  $-157.5 \pm 8.6$  &  570s, VLT/FORS2 \\
 2456804.511249  &  $-55.3 \pm 7.0$  &  570s, VLT/FORS2 \\
 2456804.565577  &  $192.3 \pm 6.1$  &  570s, VLT/FORS2 \\
 2456804.573015  &  $217.2 \pm 6.3$  &  570s, VLT/FORS2 \\
 2456804.580532  &  $235.2 \pm 6.4$  &  570s, VLT/FORS2 \\
 2456804.588222  &  $252.8 \pm 6.3$  &  570s, VLT/FORS2 \\
\hline
\end{tabular}
\end{center}
\end{table}

We have computed the radial velocities for each epoch of VLT/FORS2 and SDSS data using the same techniques described in \citet{Schreiber08} and \citet{Rebassa-Mansergas08}. In summary, we fit a double-Gaussian-line profile to the \Ion{Na}{I} absorption doublet, with rest wavelengths of 8183.27 and 8194.81 \AA, to compute a radial velocity for each epoch. The results are detailed in Table~\ref{tab:rvjour}. We then compute a periodogram of the velocities \citep{Scargle82}.

The folded radial-velocity curve of \tar\ is shown in Fig.~\ref{fig:rv}. We have computed these parameters from a best fit to the equation
\begin{equation}
v_\mathrm{r} =
K_{\rm sec}\,\sin\left[\frac{2\pi(t-t_0)}{P_\mathrm{orb}}\right]
+\gamma_{\rm sec}
\end{equation}
where $K_{\rm sec}$ is the radial velocity semi-amplitude of the companion star, $t_0$ is the time of inferior conjunction of the secondary, $P_\mathrm{orb}$ is the orbital period, and $\gamma_{\rm sec}$ is the systemic velocity of the secondary. We find $P_\mathrm{orb}=$6.89760103(60)\,hr, $\gamma_{\rm sec}=47.8\pm3.5$\,\kms, $K_{\rm sec} = 222.6\pm3.8$ \kms, and find for the inferior conjunction:
$$ t_0 \rm{(BJD_{TDB})} = 2454925.5124(12) + 0.287400043(25) \; E $$

We have assumed circular orbits, and the best-fitting solution bears this sinusoidal signature. We find that the mass function is $f=0.328\pm0.017$\,\msun. Adopting the spectroscopic masses for both components, we find that the stars are separated by $a_{\rm sep}=1.70\pm0.08$\,\rsun, and the volume-averaged secondary Roche lobe radius is roughly $RL_{\rm sec}=0.81\pm0.07$\,\rsun\ \citep{Eggleton83}. Thus, the M6 secondary should sit comfortably within its Roche lobe. The system is far from the onset of mass transfer and its evolution into a semi-detached cataclysmic variable, which will occur in roughly 13\,Gyr, when the system will have an orbital period of 1.7\,hr \citep{Schreiber03}.

\subsection{Binary Light Curve Model}
\label{sec:bin:lc}

The dominant feature of the light curve, visible by eye in Fig.~\ref{fig:lc}, is the modulation near 6.9\,hr. Taking a Fourier transform of the entire {\em K2} dataset, we find a dominant peak at $6.89756\pm0.00017$\,hr. We recognize this immediately as a less precise determination of the orbital period.

\begin{figure}
\centering{\includegraphics[width=0.995\columnwidth]{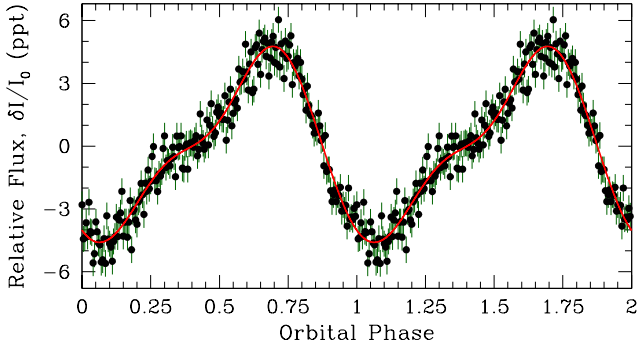}}
\caption{The light curve of our {\em K2} observations of \tar\ folded at the RV-determined orbital period of 6.8976010\,hr and repeated for clarity. We can produce a sensible model, overplotted in red, of Doppler beaming and ellipsoidal variations of the M dwarf that well describes the orbital photometric variability. \label{fig:lcfolded}}
\end{figure}

We have phased the {\em K2} photometry to the inferior conjunction defined from the radial velocities, and folded the light curve into 200 phase bins at the RV-determined orbital period. The result is displayed in Fig.~\ref{fig:lcfolded}. Both of our light curve extraction routines described in Section~\ref{sec:obs:K2} produce an identical folded light curve.

The asymmetric profile of the {\em K2} light curve can be interpreted as a combination of reprocessed light (reflection), Doppler beaming, and ellipsoidal variations of the secondary component of \tar. To confirm this, we have computed models using the {\sc lcurve} software package developed by Tom Marsh, which computes models for binary systems containing at least one white dwarf (see \citealt{Copperwheat10}). Our best-fitting model is overplotted in red in Fig.~\ref{fig:lcfolded}.

To calculate this model, we performed an initial exploration of the parameter space using a Monte Carlo analysis, with initial guesses from the atmospheric parameters described in Section~\ref{sec:spec}. This showed that the data is likely under-constrained, and that there are too many degeneracies to allow for convergence on a robust solution. We proceeded to fix essentially all parameters.

We fixed the white dwarf radius and effective temperature as well as the secondary radius to the atmospheric parameters determined in Section~\ref{sec:spec}, and used the corresponding Claret four-parameter limb-darkening coefficients of \citet{Gianninas13} and \citet{Claret11}. We convolved the best-fitting atmospheric model spectra (shown in the top panel of Fig.~\ref{fig:spec}) with the {\em Kepler} bandpass to estimate the Doppler beaming factors for each component (e.g., \citealt{Bloemen11}); we found $\left<B\right>_{\rm WD}=2.34$ and $\left<B\right>_{\rm sec}=11.69$. The beaming factor for the M dwarf is quite high, and reflects the sharp spectral features that can move in and out of the {\em Kepler} bandpass. We fixed the mass ratio at $q =  M_{\rm sec} / M_{\rm WD} = 0.326$, and assumed $K_{\rm WD} = q K_{\rm sec}$.

Our best fit has an inclination of $i=83$\,deg, for which the white dwarf contributes 61.6\,per\,cent of the flux in the {\em Kepler} light curve. Aside from inclination, we also fit for the secondary temperature (which best fit as 3030\,K) and the fraction of the irradiating flux absorbed by the companion (0.632). The fit is reasonable, with a $\chi^2_{\rm red}=1.14$.

The slight bump around phase 0.25 in the light curve corresponds to ellipsoidal variations of the M dwarf, which peak twice per orbit with roughly the same amplitude; the second maximum near phase 0.75 is enhanced due to Doppler beaming. Ellipsoidal variations are tidal distortions of the secondary and are sensitive to the mass ratio, inclination, and secondary radius (e.g., \citealt{Morris93}). Unfortunately, our models are under-constrained without more information about the component masses (especially the primary radial velocity) or system inclination. From a lack of eclipses in the light curve, we can only constrain the inclination of the system to $i<85$\,deg. If we take the component masses from Section~\ref{sec:spec} at face value, the inclination must fall between $61 < i < 85$\,deg given the mass function.

Doppler beaming imprints a modulation in the flux of a stellar binary as the source is approaching or receding, directly proportional to the radial-velocity semi-amplitude of the source as well as the slope of its spectrum through the observed filter \citep{Loeb03,Zucker07}. The dominant source of Doppler beaming in \tar\ is the secondary, hence the peak nearest phase 0.75. The peak occurs slightly before phase 0.75 due to a small amount of reprocessed flux (reflection) from the much hotter primary.

\begin{figure*}
\centering{\includegraphics[width=0.945\textwidth]{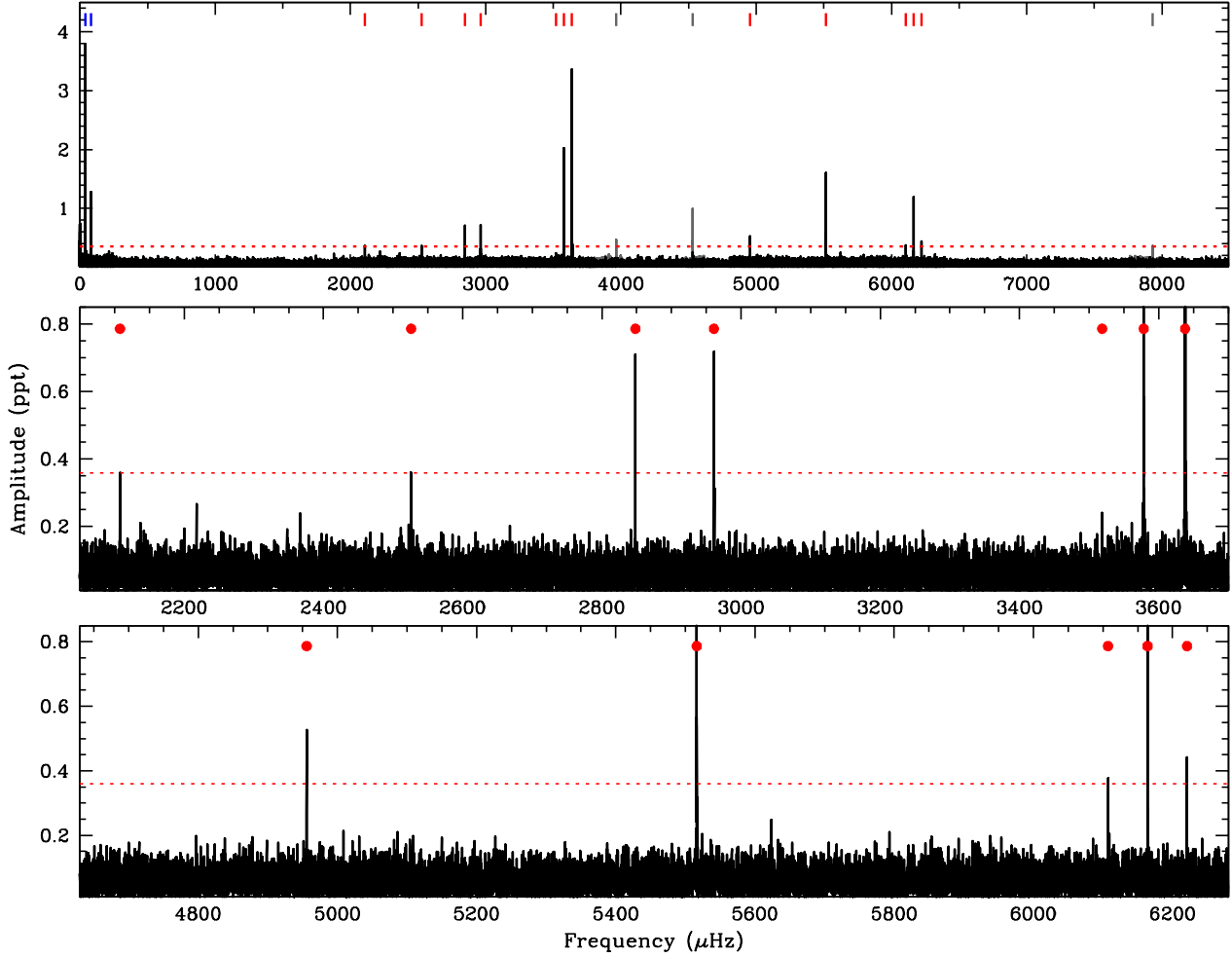}}
\caption{Fourier transforms of our {\em K2} Campaign 1 observations of \tar. The top panel shows the FT of the entire dataset out to the Nyquist frequency. The middle and bottom panels show the independent pulsation modes in more detail, marked with red dots; the full frequency solution is presented in Table~\ref{tab:J1136freq}. The dashed red line marks the significance threshold, described in the text. \label{fig:ftall}}
\end{figure*}

For now, we present our model as able to reproduce the data in hand, but do not yet consider it a robust, final solution. We have used our model to show that, with a fixed orbital period, the $t_0$ from our raw, unfolded light curve is 2454925.5145(11) \bjd, and agrees within 1$\sigma$ with the inferior conjunction defined from the spectroscopy in Section~\ref{sec:bin:rv}.

\section{Pulsation Analysis}
\label{sec:pulsations}

A Fourier transform (FT) of the 77.87\,d {\em K2} Campaign 1 light curve of \tar\ is shown in Fig.~\ref{fig:ftall}. Table~\ref{tab:J1136freq} details all of the significant periodicities detected in the light curve.

\subsection{Preliminary Mode Identifications}
\label{sec:pulsations:modeid}

\begin{table}
 \centering
  \caption{Significant frequencies in the {\em K2} Field 1 photometry of SDSS~J113655.17+040952.6 \label{tab:J1136freq}}
  \begin{tabular}{@{}lrrrr@{}}
  \hline
  ID & Frequency & Period & Amplitude & Phase \\
  & (\muhz) & (s) & (ppt) & (cycle) \\
 \hline
$f_{1,+}$ & 3638.2374 & $274.858366(88)$ & $3.493(50)$ & 0.1179(23)  \\
$f_{1,0}$ & 3578.5472 & $279.44301(14)$  & $2.272(50)$ & 0.0547(35) \\
$f_{2}$   & 5516.2411 & $181.282868(73)$ & $1.841(50)$ & 0.4957(43) \\
$f_{3,0}$ & 6164.0456 & $162.231117(89)$ & $1.213(50)$ & 0.1201(65) \\
$f_{4,+}$ & 2961.1013 & $337.71219(60)$ & $0.775(50)$ & 0.430(10) \\
$f_{4,-}$ & 2848.1735 & $351.10220(70)$  & $0.715(50)$ & 0.885(11) \\
$f_{5}$   & 4955.8397 & $201.78215(32)$  & $0.519(50)$ & 0.756(15) \\
$f_{3,+}$ & 6220.7698 & $160.75181(24)$  & $0.448(50)$ & 0.641(18) \\
$f_{6}$   & 2107.681  & $474.4551(23)$   & $0.398(50)$ & 0.145(20) \\
$f_{3,-}$ & 6107.362  & $163.73682(27)$  & $0.399(50)$ & 0.515(20) \\
$f_{7}$   & 2525.804  & $395.9135(17)$   & $0.373(50)$ & 0.767(21) \\
$f_{1,-}$ & 3518.907  & $284.1791(14)$   & $0.240(50)$ & 0.259(33) \\
\hline
$f_{\rm orb}$  & 40.2719 & $24\,831.21(62)$ & $4.063(50)$ &  \\
$2f_{\rm orb}$ & 80.5426 & $12\,415.79(47)$   & $1.347(50)$ &  \\
\hline
$8f_{\rm LC}$  & 4531.8157 & $220.66210(20)$ & $0.994(50)$ &   \\
$7f_{\rm LC}$  & 3965.3057 & $252.18737(56)$  & $0.467(50)$ &  \\
$11f_{\rm LC}$ & 7930.6581 & $126.09294(18)$  & $0.354(50)$ &  \\
\hline
\end{tabular}
\end{table}

The two significant peaks at lowest frequencies marked in blue in Fig.~\ref{fig:ftall} fall at $40.2719\pm0.0010$\,\muhz\ and its first harmonic of $80.5426\pm0.0030$\,\muhz. These peaks correspond to the orbital frequency ($f_{\rm orb}$) and its first harmonic. Our light curve fits discussed in Section~\ref{sec:bin:lc} reveal far more information about the binary parameters, so we will not discuss these signals further.

Short-cadence {\em Kepler} data is contaminated by spurious frequencies arising from harmonics of the long-cadence exposure length ($f_{\rm LC}=566.479$\,\muhz). We detect three such artifacts with significance in our FT ($7f_{\rm LC}$, $8f_{\rm LC}$, and $11f_{\rm LC}$), and mark them in light gray in Fig.~\ref{fig:ftall}. We do not include these signals in our analysis.

The pulsations of the white dwarf in \tar\ are confined to a region between $162.2-474.5$\,s, with a weighted mean period of 263.3\,s \citep{Clemens93}. Such relatively short pulsation periods are entirely consistent with the spectroscopically determined temperature of $12\,330$\,K (see \citealt{Mukadam06}); hot DAVs near the blue edge of the instability strip have shallower convection zones and thus drive shorter-period pulsations than do their cooler counterparts \citep{Goldreich99}.

\begin{figure*}
\centering{\includegraphics[width=0.498\textwidth]{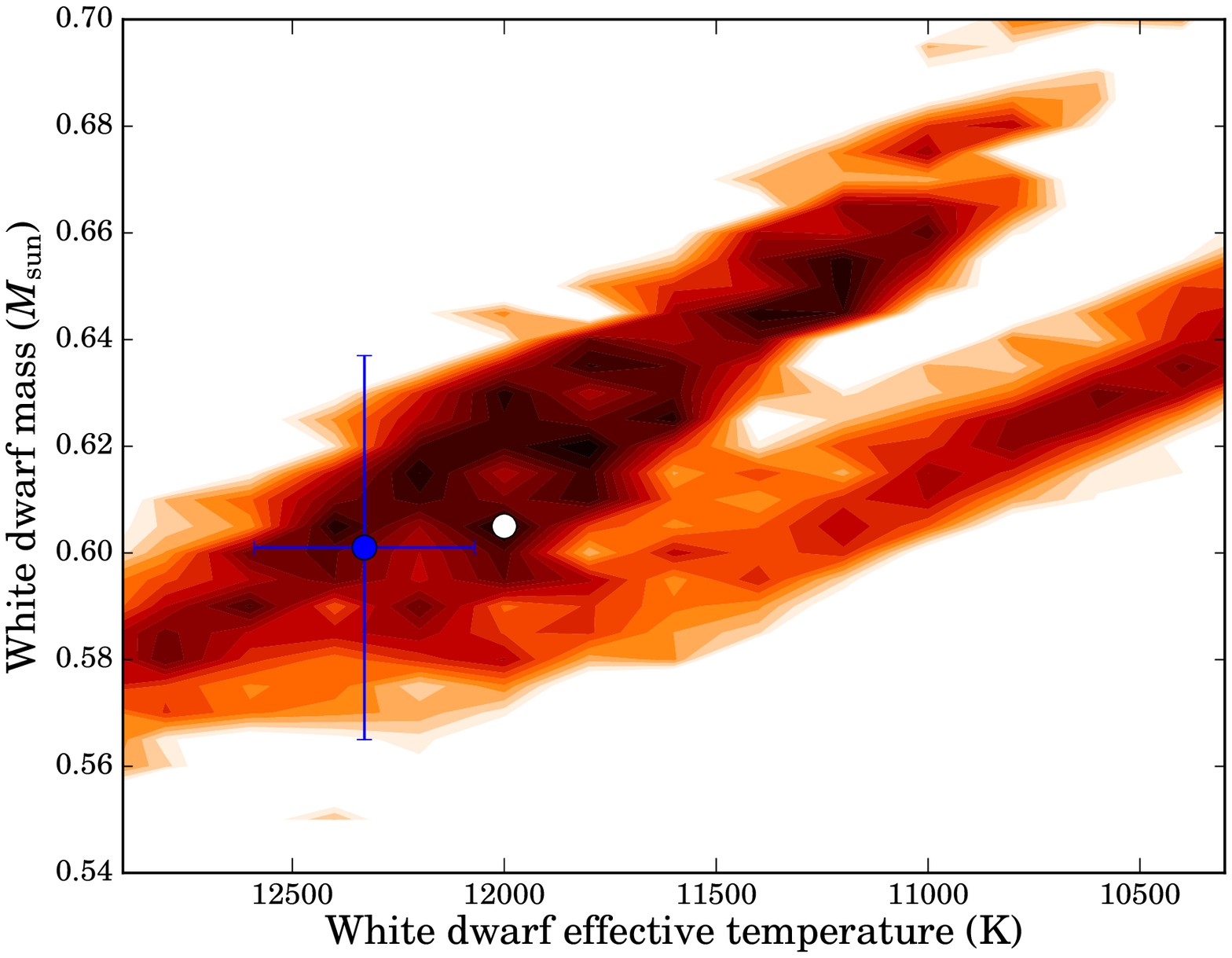}\includegraphics[width=0.46278\textwidth]{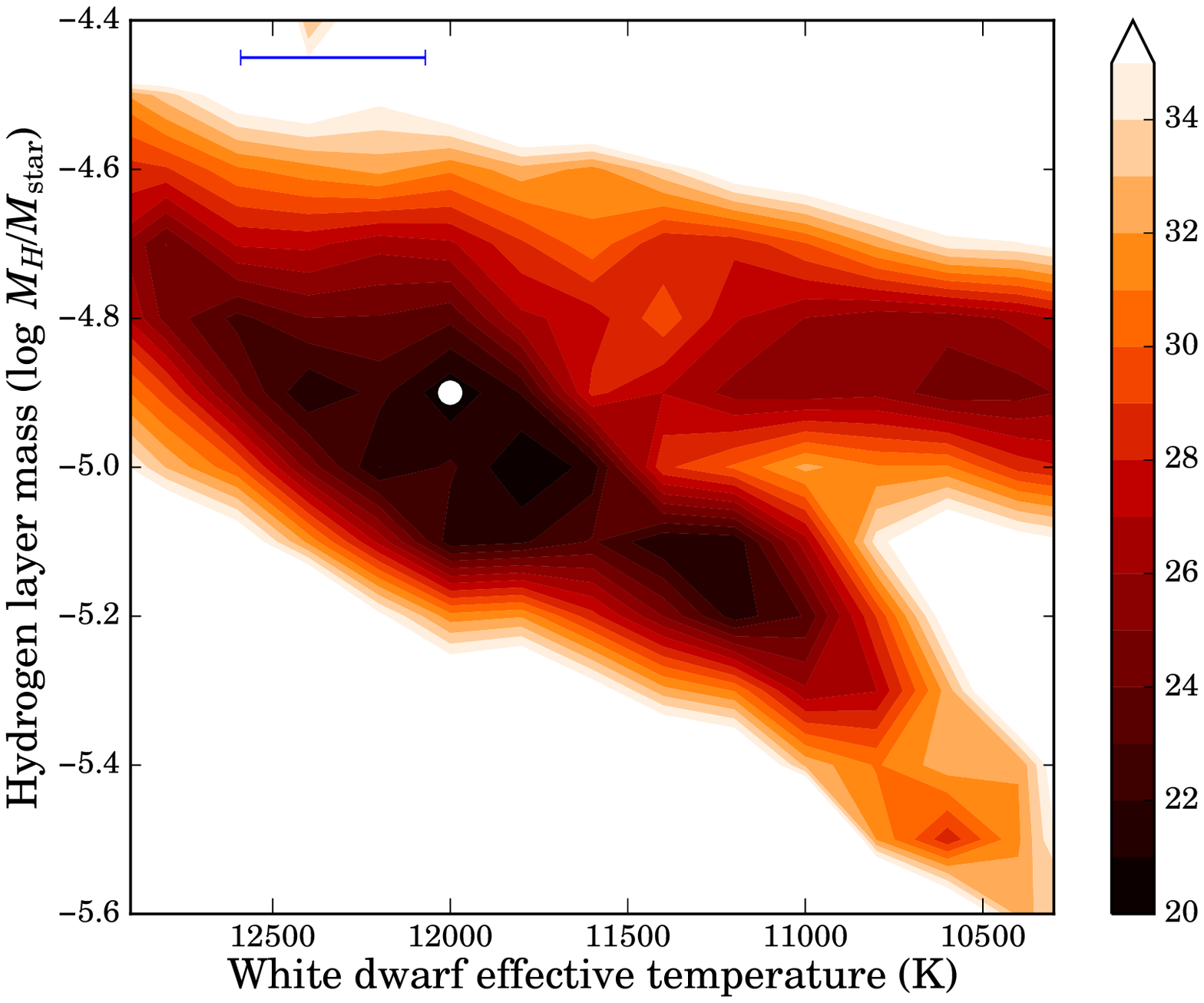}}
\caption{Contours showing the goodness of fit, $\sigma_{\rm RMS}$, from our asteroseismic analysis of \tar. The scale runs linearly in tenths of seconds; darker regions correspond to a better fit. The best asteroseismic solution is marked as a white dot and the 1$\sigma$ spectroscopically determined atmospheric parameters are marked in blue. This is a five-dimensional parameter space, so to make these two-dimensional projections the best-fitting values for the other parameters were optimized at each grid point. \label{fig:seis}}
\end{figure*}

We have computed a significance threshold following the method outlined in \citet{Greiss14}. In short, we have randomly shuffled the fluxes for each point in the light curve, and computed the highest peak in the entire resultant FT for $10\,000$ random permutations. The 2$\sigma$ and 3$\sigma$ lines are defined by the lines for which 95.5 and 99.7 per\,cent of the highest overall amplitude recorded in each shuffled light curves fall below that threshold. They are 0.330 and 0.359 ppt, respectively. The threshold we adopt is conservative since the basis light curve includes significant signals, notably the orbital modulation. However, no questionable signals fall between our 2$\sigma$ and 3$\sigma$ lines, so we adopt the 99.7 per\,cent threshold (0.359 ppt) for simplicity.

We identify 16 signals in the FT above the significance threshold, which we describe in Table~\ref{tab:J1136freq}. The quoted period, amplitude, and phase uncertainties in Table~\ref{tab:J1136freq} come from a simultaneous least-squares fit to the data using the software package {\sc Period04} \citep{Lenz05}.

All signals cleanly prewhiten from the light curve. The only frequency with a marginally significant residual occurs at 2961.585\muhz\ (with $0.341\pm0.050$\,ppt amplitude). We have also prewhitened the light curve by three very low frequencies, at 0.11842\,\muhz, 0.6624\,\muhz, and 0.8118\,\muhz, respectively. We do not believe these signals are astrophysical in nature, but are rather long-term trends that have not been perfectly removed in the light curve extraction.

We note that since the white dwarf is contributing roughly 62\,per\,cent of the flux in the {\em Kepler} bandpass, the pulsation amplitudes are all underestimated by a factor of roughly 1.61. The amplitudes have not been adjusted in Table~\ref{tab:J1136freq}, but this should be taken into account when comparing to pulsation amplitudes observed at other wavelengths.

In order to perform an asteroseismic analysis, we must know the spherical degree ($\ell$) of the pulsation mode in question. We begin with the initial assumption that all modes are the two lowest degrees possible ($\ell=1,2$) out of simplicity and because $\ell\geq3$ modes suffer from significant geometric cancellation \citep{Dziembowski77}. Additionally, we note that three regions of the FT have signals separated in frequency by a similar amount, roughly $57$\,\muhz. We assume that these are rotationally split multiplets (see \citealt{Dolez81}), and discuss these signals further in Section~\ref{sec:pulsations:rotation}.

In Table~\ref{tab:J1136freq} we have identified seven independent pulsation modes, three of which we identify as having more than one rotationally split component. We have tentatively identified the central ($0$), prograde ($+$), and retrograde ($-$) components of these three modes. In one case, $f_{1,-}$, we have used the expectation that this may be a rotationally split component of a triplet to relax the significance threshold, and include this low-amplitude mode in our analysis. 

All of the pulsation signals we have identified in Table~\ref{tab:J1136freq} are extremely stable in both amplitude and phase. Such frequency stability is a feature of hotter DAVs, which show extremely stable pulsation modes (e.g., G117-B15A, \citealt{Kepler05}). In fact, {\em every} mode in \tar\ shows exceptional amplitude and phase stability, such that if we break the {\em K2} data into four $\sim$3-week subsets, a linear least-squares fit for all values are consistent within the statistical uncertainties.

\subsection{Preliminary Asteroseismic Analysis}
\label{sec:pulsations:seism}

We have used the seven independent pulsation modes identified from the {\em K2} Campaign 1 light curve as input for an initial asteroseismic analysis. Our asteroseismic analysis proceeded in the same way as described in more detail in \citet{BK08}.

In short, we matched the observed pulsation periods ($P_{\rm obs}$) to theoretical periods ($P_{\rm calc}$) computed from adiabatically pulsating a grid of models derived from an updated version of the White Dwarf Evolution Code \citep{LvH75}. Our core composition profiles follow the smooth models of \citet{Salaris97}, and feature a sharp transition between the carbon and helium layer, where diffusive equilibrium is assumed for this transition and no diffusion coeffecients are used.

The models had five variable parameters: effective temperature, overall white dwarf mass, the mass of the hydrogen envelope ($M_{\rm H}$), the central oxygen abundance relative to carbon ($X_{\rm O}$), and the location of the edge of the homogeneous carbon-oxygen core ($X_{\rm fm}$). Our model grids were spaced in steps of 200\,K, 0.005\,\msun, 0.10\,dex, 0.10\,dex, and 0.05\,dex, respectively, with a range of $10\,000-13\,000$\,K, $0.55-0.80$\,\msun, $10^{-4.0}-10^{-6.0}$, 0.50-0.99\,dex, and 0.1-0.8\,dex, for a total of 1\,541\,202 models. The mass of the helium envelope ($M_{\rm He}$) was fixed at $M_{\rm He}/$\mstar$=10^{-2.0}$.

We assume the rotationally split modes with three components correspond to a pulsational triplet, and thus identify $f_1$, $f_3$, and $f_4$ as $\ell=1$ modes. We make no assumption about the identification of the other modes, and minimized the overall residual $\sigma_{\rm RMS}$:
\begin{equation}
\sigma_{\rm RMS} = \sqrt{\frac{\sum(P_{\rm calc}-P_{\rm obs})^2}{n_{\rm obs}}}
\end{equation}
where $n_{\rm obs}=7$ is the number of observed modes.

\begin{table}
 \centering
  \caption{Summary of the observed periods and the best-fitting asteroseismic model. We have fixed the $\ell$ marked by a $^{*}$ symbol. \label{tab:seis}}
  \begin{tabular}{@{}lrrrrr@{}}
  \hline
  ID & Observed & Model & $\ell$ & $k$ & $C_{k,\ell}$  \\
   & $P_{\rm obs}$ (s) & $P_{\rm calc}$ (s) &  &  &   \\
 \hline
$f_{1}$ & $279.44$ & $280.78$ & 1$^{*}$ & 3  & 0.350  \\
$f_{2}$ & $181.28$ & $183.41$ & 2       & 4  & 0.054  \\
$f_{3}$ & $162.23$ & $164.77$ & 1$^{*}$ & 1  & 0.490  \\
$f_{4}$ & $344.28$ & $345.15$ & 1$^{*}$ & 5  & 0.493  \\
$f_{5}$ & $201.78$ & $199.73$ & 2       & 5  & 0.157  \\
$f_{6}$ & $474.46$ & $471.19$ & 1       & 7  & 0.412  \\
$f_{7}$ & $395.91$ & $396.23$ & 2       & 12 & 0.113  \\
\hline
\end{tabular}
\end{table}

The best-fitting asteroseismic solution is detailed in Table~\ref{tab:seis}, with a $\sigma_{\rm RMS}=2.02$\,s. It arose from the following parameters: \teff$=12000$\,K, $M_{\rm WD}=0.605$\,\msun, $M_{\rm H}/$\mstar$=10^{-4.9}$, $X_{\rm O}=0.99$, and $X_{\rm fm}=0.70$. The top 20 best fits all shared the same $\ell$ identifications, with $2.09 < \sigma_{\rm RMS} < 2.47$\,s: their means and standard deviations form an asteroseismic temperature ($12260\pm270$\,K) and mass estimate ($0.605\pm0.014$\,\msun). The other parameters were consistent within 0.1 dex of the best-fitting asteroseismic solution. The shaded contours in Fig.~\ref{fig:seis} show only solutions with $\sigma_{\rm RMS}<3.5$\,s, and  represent just 0.2\,per\,cent of all models.

The best-fitting models consistently prefer a high oxygen-carbon ratio in the core, but expectations of this parameter are fairly unconstrained (see the discussion in \citealt{BK11}), and $X_{\rm O}$ has a relatively small impact on DAV period spectra (see Table~5 in \citealt{BK08}). The models of \citealt{Althaus10b} predict $X_{\rm O}\sim0.7$ for a 0.6\,\msun\ white dwarf.

Given how precisely we were able to determine the pulsations periods with our {\em K2} light curve (to an average of 0.00066\,s), the best-fitting asteroseismic model has a rather large $\sigma_{\rm RMS}$ of 2.02\,s. This indicates that our best fit may not have fully converged on a final solution, or that the models inherently lack some physical details. However, we are encouraged by the good agreement between the atmospheric parameters (temperature and mass) determined by both asteroseismology and spectroscopy. Our preliminary asteroseismic analysis also improves our estimate of the white dwarf rotation period.

\subsection{White Dwarf Rotation Period}
\label{sec:pulsations:rotation}

\begin{figure}
\centering{\includegraphics[width=0.93\columnwidth]{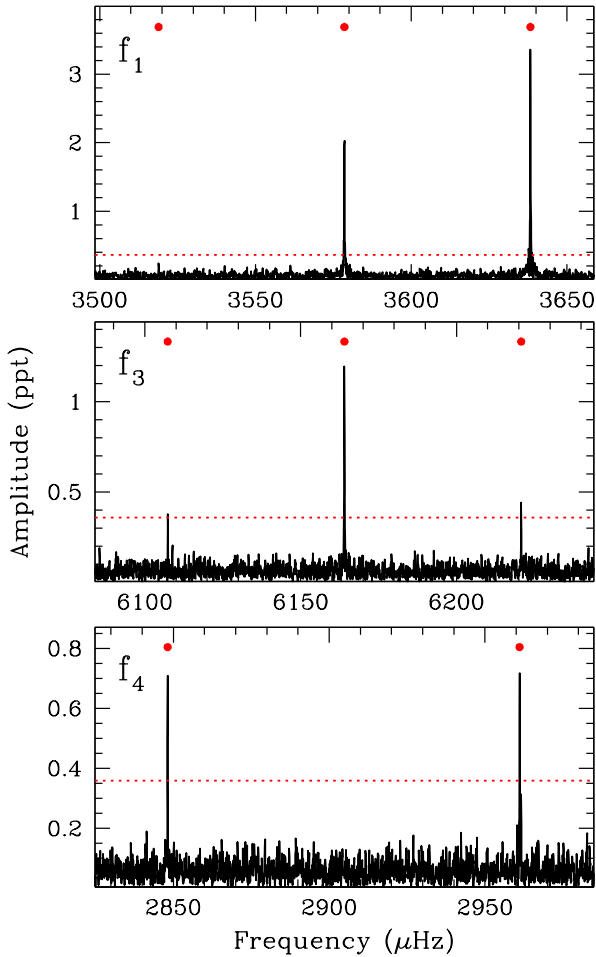}}
\caption{Detailed regions of the Fourier transform of \tar\ showing the three different rotationally split pulsation modes, all viewed with the same frequency scale. The triplets are split by roughly 57\,\muhz, which corresponds to a rotation period of roughly 2.49\,hr (see Section~\ref{sec:pulsations:rotation}). Our interpretation suggests that the central ($m=0$) component of $f_4$ is not observed. \label{fig:fttrips}}
\end{figure}

The symmetric frequency spacing of the $f_1$, $f_3$, and $f_4$ modes, illustrated in Fig.~\ref{fig:fttrips}, suggests that these are rotationally split multiplets of three independent pulsations modes. Since we see no evidence that these modes have five components, we assume that these are rotationally split triplets of three $\ell=1$ multiplets, each of a different radial overtone number ($k$).

To first order, the frequency splitting is related to the rotation of the white dwarf by the relation
\begin{equation}
\delta \nu = m (1 - C_{k,\ell}) \Omega
\end{equation}
where $\Omega$ is the rotation frequency of the star and the $C_{k,\ell}$ value represents the moment of inertia of the mode \citep{Unno89}. In the asymptotic limit, $C_{k,\ell}\approx1/\ell(\ell+1)$. However, this does not hold for trapped modes, or modes of low radial order ($k<10$), which is very likely the case for the relatively hot DAV in \tar. Fortunately, our best-fitting asteroseismic model has predictions for these $C_{k,\ell}$ values, which we have listed in Table~\ref{tab:seis}. Notably, $f_1$ has a very strong deviation from the asymptotic value (0.5).

We find the frequency splitting for each of the three triplets: $\delta f_1 = 59.6890(29)$\,\muhz, $\delta f_3 = 56.7057(92)$\,\muhz, and $\delta f_4 = 56.464(11)$\,\muhz. For each triplet, we computed the spacing between the central component and the prograde or retrograde component, and we average these values weighted by their uncertainties. For $f_4$ we have simply taken half the frequency difference between $f_{4,-}$ and $f_{4,+}$. Using the $C_{k,\ell}$ of our best asteroseismic model, the observed frequency splittings correspond to rotation periods of 3.02\,hr, 2.50\,hr, and 2.49\,hr.

Given the excellent agreement for $f_3$ and $f_4$, we do not use the trapped mode $f_1$ in our estimate of the overall rotation period but rather use it to constrain the uncertainties in our determination. We conclude that this white dwarf is rotating at $2.49\pm0.53$\,hr.

One troubling aspect of the three rotationally split triplets is the inconsistent relative amplitudes: the central component of $f_3$ has the highest amplitude of the multiplet, whereas the central component of $f_4$ is not detected. The central component of $f_1$ has the second-highest amplitude, while the prograde component appears strongest. If the system is at high inclination, as we found in Section~\ref{sec:bin:lc}, geometric considerations predict the amplitude of $m=\pm1$ are much larger than $m=0$ for $\ell=1$ modes, and that $m=\pm2,0$ are favoured for $\ell=2$ modes (see \citealt{Pesnell85}). The expected amplitude ratios work for $f_4$, but not the other two multiplets. We note that if the multiplets are all $m=\pm2,0$ components of $\ell=2$ modes rather than $m=\pm1,0$ components of $\ell=1$ modes, the splittings would correspond to a rotation period of $8.3\pm1.8$\,hr using the asymptotic relation for $\ell = 1,2$ modes.

Stochastically excited stars, such as solar-like oscillators, have equipartition of energy, such that the amplitudes of all components of a rotationally split multiplet are symmetric and predictable \citep{GS03}. However, observationally, this does not appear to be the case for pulsating white dwarfs: different multiplets in the same star almost never have the expected relative amplitudes for a given inclination and the same intrinsic mode amplitudes. For example, the amplitudes of both the $\ell=1$ and the $\ell=2$ modes of one of the best-studied pulsating white dwarfs, PG1159$-$035, show inconsistent relative amplitudes \citep{Winget91}. Similarly, the amplitude ratios of another class of pulsating compact objects, hot subdwarfs, do not show consistent relative amplitudes for all $\ell=1$ modes of the same star \citep{Pablo12}. Thus, the different relative amplitudes of the observed rotationally split multiplets in \tar\ is not problematic to our analysis.

We have searched the FT for a signal at the corresponding spin frequency, $f_{\rm spin}=111.6$\,\muhz, but do not detect any peaks in this region above 0.2\,ppt. We do not see a photometric signal of the white dwarf rotation period.

Finally, any slight asymmetry in what should be the even frequency splitting of a pulsation multiplet can constrain the presence of a weak magnetic field, which also acts to break spherical symmetry and cause subtle changes in the frequency spacing \citep{Jones89}. We find that the two triplets with all three components, $f_1$ and $f_3$, are symmetric within the uncertainties except for the prograde component of $f_1$, which is still symmetric within 2$\sigma$. Thus, we can only put an upper limit on the magnetic field strength. Following the method and the scaling factor detailed in \citet{Winget94}, the frequency splitting symmetry to better than 0.023\,\muhz\ roughly constrains the dipolar magnetic field of the white dwarf in \tar\ to $<10$\,kG. Such a low global magnetic field may be consistent with the lack of a corresponding photometric signal of the white dwarf spin.

\section{Discussion}
\label{sec:discussion}

The pulsating component in \tar\ is the first carbon-oxygen core white dwarf known to pulsate and also have a close, detached companion \citep{Pyrzas15}. The only other pulsating white dwarfs with close companions (which have also evolved through at least one and probably two common-envelope phases) are the extremely low-mass ($M_{\rm WD}<0.25$\,\msun) white dwarfs showing pulsations \citep{Hermes12,Hermes13a,Hermes13,Kilic15}, as well as the no-longer-detached pulsating white dwarfs in cataclysmic variables (e.g., \citealt{vZ04}).

As discussed in Section~\ref{sec:pulsations:seism}, the seven independent pulsation modes in \tar\ are exceptionally well determined, and provide the possibility for the first in-depth analysis of the interior structural effects of common-envelope evolution. Unfortunately, our asteroseismic analysis is hampered by the fact that there are nearly as many free parameters in our models as there are observed pulsation modes. We have shown that the 20 best-fitting asteroseismic models are in good agreement with the spectroscopically determined atmospheric parameters. However, the best fit is not excellent: it has a $\sigma_{\rm RMS}=2.02$\,s, which is significantly higher than the uncertainty for each pulsation mode ($\sigma<0.001$\,s).

For these preliminary asteroseismic fits, we have consistently found a hydrogen layer mass, which controls the rate at which the white dwarf cools, that is thinner than we expect for an isolated white dwarf. Our best fit finds this layer has a mass of $M_{\rm H}/$\mstar$\approx10^{-4.9}$, thinner than the canonical value for DA white dwarfs, $M_{\rm H}/$\mstar$=10^{-4.0}$ (e.g., \citealt{Althaus05}). The seismologically derived hydrogen layer mass is nearly an order of magnitude smaller than expected from shell burning in the evolutionary models, and could indicate that some of the hydrogen mass was stripped in the common envelope phase. However, the fact that we have not considered thinner helium-layer masses and fixed that parameter at the maximal value of $M_{\rm He}/$\mstar$\approx10^{-2.0}$ may have an influence on the derived hydrogen-layer mass.

The only previous constraints on the hydrogen-layer mass of a PCEB have required comparing precision studies of deeply eclipsing systems to model-dependent mass-radius relationships. Such studies have shown that the white dwarfs in PCEBs contain both a relatively thick hydrogen-layer mass, $M_{\rm H}/$\mstar$\approx10^{-4}$, as in the case of the 0.535\,\msun\ NN~Ser \citep{Parsons10}, as well as a relatively thin hydrogen-layer mass, $M_{\rm H}/$\mstar$\approx10^{-8}$, as in the case of the 0.439\,\msun\ SDSS J1212$-$0123 \citep{Parsons12}.

Pulsations also have the potential to further refine the binary parameters of this unique system. As mentioned in Section~\ref{sec:pulsations:modeid}, the pulsations visible in \tar\ are notably stable, and a single frequency can well-describe each peak in the FT. However, we know that the phases of the pulsations are periodically changing at the orbital period. This is a light-travel-time effect as the white dwarf moves around its centre of mass (e.g., \citealt{Barlow11}).

Constraining this phase variation would allow us to directly measure $K_{\rm WD}$ and provide a model-independent solution to the physical parameters. Given the mass estimates from Section~\ref{sec:spec}, we expect the amplitude of this phase variation to be $\tau=1.8\pm0.7$\,s. However, the {\em K2} light curve is simply too noisy to enable our directly measuring this quantity. We also attempted, unsuccessfully, to directly detect sidelobes in the amplitude spectrum, manifest as splittings from the central components of our pulsations set apart by the orbital frequency. We expect the sidelobes to have roughly 2\,per\,cent the amplitude of the central component, which is simply buried in the noise of the FT \citep{Shibahashi12}. We can only set an upper limit of $\tau<10$\,s on this phase variation from the {\em K2} data.

Our radial-velocity observations were carried out to exclusively cover the \Ion{Na}{I} absorption doublet at wavelengths longer than 7500\,\AA, so we have no spectroscopic information on the white dwarf radial velocities. Measuring the white dwarf radial-velocity semi-amplitude directly from the Balmer lines should be possible, since in all likelihood $K_{\rm WD}>70$\,\kms. This will fully solve the orbital parameters, yielding a precise estimate for the mass ratio, as well as a gravitational redshift for the white dwarf.

Despite our inability to use the oscillation phases to constrain the binary mass ratio, the exceptional stability of the pulsations can be monitored longer-term to watch the orbital period evolution of the system. Roughly 90\,per\,cent of detached, eclipsing PCEBs observed for more than 5\,yr show significant changes in the mid-eclipse times inconsistent with secular evolution (for a list, see \citealt{Zorotovic13}). Some of these eclipse variations have been attributed to light-travel-time effects caused by circumbinary planets (e.g., \citealt{Marsh14}). However, follow-up studies have been unable to directly detect these external companions and have been damaging to the circumbinary interpretation of eclipse timing variations \citep{Hardy15}.

With follow-up photometry it will be possible to use the stable pulsations in \tar\ as a clock to monitor for external light-travel-time effects. Modulations in the pulse arrival times must come from light-travel-time effects, whereas mid-eclipse-time variations can arise from orbital period variations caused by other means, such as Applegate's mechanism \citep{Applegate92}. To enable further monitoring, we have included in Table~\ref{tab:J1136freq} the phase point for the time-of-maximum of each pulsation mode from the {\em K2} observations. These phases align with the mid-point of the {\em K2} light curve, 2456851.5625920 \bjd.

Our {\em K2} observations of the white dwarf in \tar\ provide one of the most precise constraints on the rotation period of an object that has gone through common-envelope evolution {\em and is still fully detached}. The only other well-constrained rotation period of a PCEB has been determined for the eclipsing system V471 Tau, which has a 9.25-min spin period, although the K2V companion is nearly Roche-lobe filling and may have previously transferred mass and angular momentum onto the white dwarf (see \citealt{Hussain06} and \citealt{OBrien01}).

In the context of other asteroseismically deduced rotation periods of white dwarfs, \tar\ is rotating more rapidly than any other known isolated white dwarf with clearly resolved pulsation multiplets \citep{Kawaler04,Kawaler14}. The previous record-holder is the hot pre-white-dwarf PG~2131+066, which rotates with a period of 5.07\,hr; notably, it is also in a wide binary with a K7V companion, although the two are currently separated by at least 200\,au and so did not undergo a common-envelope event \citep{Kawaler95,Reed00}.

Over time, gravitational radiation and magnetic braking from the M dwarf will bring the two components of \tar\ closer together, and the system will likely become a cataclysmic variable. Interestingly, all non-magnetic white dwarfs in cataclysmic variables have rotation periods on the order of minutes or faster, much more rapid than observed for the white dwarf in \tar\ (e.g., \citealt{Sion94,Sion95,Szkody02}). Over time, mass transfer in cataclysmic variables is expected to transport enough angular momentum to spin up the white dwarf to near break-up velocities (e.g., \citealt{King91}).

For now, the components of \tar\ are detached and still well separated, and the secondary fills roughly $24\pm20$\,per\,cent of its Roche lobe. It is therefore very unlikely that episodes of stable mass transfer occurred in the past that could have considerably affected the current white dwarf rotation. Therefore, the measured rotation period ($2.49\pm0.53$\,hr) and the upper limit on the magnetic field ($<10$\,kG) put interesting constraints on angular momentum transport in evolved stars. Notably, angular momentum transport in giants is still an open problem in astrophysics \citep{Cantiello14}. First constraints on the problem of core spin-up and envelope coupling in red giants are being explored using seismic data on first-ascent giants from the original {\em Kepler} mission \citep{Beck12,Mosser12}.

If isolated white dwarf rotation periods reflect some sort of coupling with the red-giant envelope, then the relatively fast rotation of the white dwarf in \tar\ could have resulted from truncated red-giant evolution via common-envelope ejection. Additionally, the relatively low magnetic field strength of the white dwarf in \tar\ may have precluded the development of magnetic torques to slow the white dwarf rotation after the common-envelope event \citep{Suijs08}.

Finally, we have shown that the white dwarf rotation period is not synchronized with the orbital period of roughly $6.9$\,hr. Similar results from three hot subdwarfs with M dwarf companions in the original {\em Kepler} mission showed a similar lack of synchronization, although those subdwarfs rotate much slower than the orbital period \citep{Pablo11,Pablo12}. Finding more such systems help constrain theoretical models for timescales of tidal synchronization.

\section{Summary and Conclusions}
\label{sec:summary}

We have used extensive ground-based spectroscopy and space-based photometry to constrain the physical parameters of the first known pulsating white dwarf that underwent a single common-envelope event, \tar. We summarize these values in Table~\ref{tab:physical}.

\begin{table}
 \centering
  \caption{Summary of system parameters determined for \tar. Parameters were derived from (1) VLT/FORS2 spectroscopy, (2) SOAR/Goodman spectroscopy, (3) SDSS spectroscopy, and (4) {\em K2} Campaign 1 light curve. \label{tab:physical}}
  \begin{tabular}{@{}llr@{}}
  \hline
  Parameter & Value & Source  \\
 \hline
Orbital period, $P_{\rm orb}$ & 6.89760103(60)\,hr & (1) \\
WD temperature (spectroscopy) & 12\,330(260)\,K      & (2) \\
WD temperature (seismology)   & 12\,260(270)\,K      & (4) \\
WD \logg\                     & 7.99(06)\,cgs      & (2) \\
WD mass (spectroscopy)        & 0.601(13)\,\msun\  & (2) \\
WD mass (seismology)          & 0.605(14)\,\msun\  & (4) \\
WD radius (seismology)        & 0.013\,\rsun\      & (4) \\
Secondary spectral type       & M$6\pm1$V           & (3) \\
Secondary radius              & 0.195(90)\,\rsun\  & (3) \\
Secondary mass                & 0.196(85)\,\msun\  & (3) \\
WD distance                   & 125(07)\,pc        & (3) \\
Secondary distance            & 121(56)\,pc        & (3) \\
Secondary RV, $K_{\rm sec}$   & 222.6(3.8)\,\kms\  & (1) \\
Orbital separation, $a_{\rm sep}$ & 1.70(08)\,\rsun\ & (1,2,3) \\
Secondary roche lobe          & 0.81(07)\,\rsun\   & (1,2,3) \\
WD rotation period              & 2.49(53)\,hr       & (4) \\
WD magnetic field             & $<10$\,kG          & (4) \\
WD hydrogen layer mass        & $M_{\rm H}/$\mstar$\approx10^{-4.9}$ & (4) \\
\hline
\end{tabular}
\end{table}

Both our time-series spectroscopy and photometry vary at 6.89760103\,hr, which we determine is the orbital period of this evolved binary. Our photometry from the extended {\em Kepler} mission shows orbital variability that can be well-reproduced with a model of Doppler beaming, reflection, and ellipsoidal variations of the M6V main-sequence companion. We have matched the seven observed independent pulsation modes to theoretical evolution models, and find reasonable agreement between the spectroscopically determined white dwarf temperature and mass. Asteroseismology also allows us to peer into the interior of the white dwarf and constrain its hydrogen layer mass to $M_{\rm H}/$\mstar$\approx10^{-4.9}$, although we only consider this a preliminary asteroseismic investigation and urge caution in over-interpreting this result.

Notably, the observed splittings of the pulsations demonstrate that the white dwarf in this close binary is rotating at a period of $2.49\pm0.53$\,hr, more rapidly than the orbital period. These splittings are symmetric within the uncertainties, which sets an upper limit on the dipole magnetic field of the white dwarf to $<10$\,kG. Both values are extremely useful inputs for further modeling of the previous common-envelope event that shaped the present binary configuration of \tar, as well as angular momentum transport in stars that have likely underwent truncated red-giant evolution.

\section*{Acknowledgments}

We thank the anonymous referee, whose thorough comments significantly improved this manuscript. We wish to acknowledge fruitful discussions with Jim Fuller, Danny Steeghs, and Roberto Raddi. J.J.H., B.T.G., and P.C. acknowledge funding from the European Research Council under the European Union's Seventh Framework Programme (FP/2007-2013) / ERC Grant Agreement n. 320964 (WDTracer). J.T.F. acknowledges support from the NSF under award AST-1413001. M.H.M. and D.E.W. gratefully acknowledge the support of the NSF under grant AST-1312983. M.H.M. acknowledges the support of NASA under grant NNX12AC96G. T.R.M. was supported under a grant from the UK's Science and Technology Facilities Council (STFC), ST/L000733/1. A.G. acknowledges support provided by NASA through grant number HST-GO-13319.01 from the Space Telescope Science Institute, which is operated by AURA, Inc., under NASA contract NAS 5-26555. M.R.S. acknowledges support from FONDECYT (grant 1141269) and from the Millenium Nucleus RC130007 (Chilean Ministry of Economy). A.R.M. acknowledges financial support from the Postdoctoral Science Foundation of China (grants 2013M530470 and 2014T70010) and from the Research Fund for International Young Scientists by the National Natural Science Foundation of China (grant 11350110496).

This work is based on observations collected at: the European Organisation for Astronomical Research in the Southern Hemisphere, Chile (083.D-0862, 093.D-0300), the SOAR telescope, and the {\em Kepler} spacecraft (GO1015). The SOAR telescope is a joint project of the Minist\'{e}rio da Ci\^{e}ncia, Tecnologia, e Inova\c{c}\~{a}o (MCTI) da Rep\'{u}blica Federativa do Brasil, the U.S. National Optical Astronomy Observatory (NOAO), the University of North Carolina at Chapel Hill (UNC), and Michigan State University (MSU). Funding for the {\em Kepler} mission is provided by the NASA Science Mission Directorate. 

{\it Facilities:} Kepler, K2, VLT, SOAR, SDSS

\label{lastpage}

\end{document}